# A Graph-based Ranking Approach to Extract Key-frames for Static Video Summarization

**Saikat Chakraborty**[1,2]

**Abstract** Video abstraction has become one of the efficient approaches to grasp the content of a video without seeing it entirely. Key frame-based static video summarization falls under this category. In this paper, we propose a graph-based approach which summarizes the video with best user satisfaction. We treated each video frame as a node of the graph and assigned a rank to each node by our proposed *VidRank* algorithm. We developed three different models of *VidRank* algorithm and performed a comparative study on those models. A comprehensive evaluation on 50 videos from open video database using objective and semi-objective measures indicates the superiority of our static video summary generation method.

## 1 Introduction

In the recent era, the availability of high end video based information and time constraints problem in daily life of user has evolved a specific research domain called *Video abstraction*. This abstract representation of information ease user to acquire knowledge of a large content in a very short time. *Video abstraction* is a mechanism for generating a short summary of a video, which can either be a sequence of stationary images (key frames) or moving images (video skims) [1].Key frames are a set of images which are the representative of the entire video. These are static in nature. Video skim is actually a set of clips of the original video. This is also called moving story board. This type of abstract is generated by segmenting the original video by identifying transitional point (e.g. cut, fade-in, fade-out, dissolve, wipe).The advantage of video skim is that it includes both audio and motion feature which increase expressiveness of the abstract. But in terms of storage optimization, static story board is more feasible than moving story board. On the other hand, since they are not restricted by any timing or synchronization issues, once key frames are extracted, there are further possibilities of organizing them for browsing and navigation purposes rather than the strict sequential display of video skims [1].

In this paper we present a graph-based approach which extract key-frames based on our proposed *VidRank* algorithm which is basically influenced by the renowned “page-rank” [2] algorithm. We have used both color and texture features to generate a hybrid feature vector which is used to more specifically identify images for retrieval purpose. We have augmented page-rank algorithm by introducing penalization process and created three different models of our proposed *VidRank* algorithm. Finally a comparative performance study is performed with three different state-of-the-art approaches [3-5], on 50 videos from the Open Video Project [14] with an evaluation framework which consist of an objective metric and one semi-objective metric. The result shows that our proposed approach is superior to some of the state-of-the-art approaches and also comparable with some of the approaches.

---

[1] S. Chakraborty was a post graduate student of Department of Computer Science & Engineering, Jadavpur University, Kolkata ,West Bengal, India (saikat.sc@gmail.com) during the work.
[2] S. Chakraborty is now a doctorate research scholar in National Institute of Technology Rourkela, Odisha, India.

The rest of this paper organized as follows: section 2 discusses related work and our contribution. Section 3 describes our proposed method. A comparative performance study presented in section 4. Finally, section 5 concludes the paper with an outline of future research direction.

# 2 Related Work

A comprehensive and comparative review of static video summarization approaches can be found in [1], [6]. But our rank-based summarization approach is less explored still now in video summarization domain, so we have not found completely related work in the literature. Instead we have found some similarities of our research work with cluster based video summarization approaches. One of the techniques that Hanjalic and Zhang [7] have proposed for static video summarization is based on an optimal clustering through cluster-validity analysis. Depending on the number of frames in video a partition clustering is applied several times. But partition clustering makes the above process computationally expensive. Xinding *et al.*[15] has extracted pre-defined fixed number of frames as key frames by content-based adaptive clustering. But because it only considers histogram to extract color feature, it suffers from spatiality problem, also for the videos where there is no so much content change occur this method tends to generate redundant key frames. Furini *et al*.[5] proposed STIMO (STIll and MOving Video Storyboard); a model for video summarization which is based on modified Furthest-Point-First (FPF) algorithm. After fixed rate pre-sampling a color histogram-based feature vector is constructed and then based on FPF algorithm distinct clusters are generated. Generalized Jacquard Distance (GJD) is used to measure dissimilarity between consecutive frames. This method suffers from GJD based dissimilarity measure which affects the content representation of the key frame set. In [3] Avila *et al.* presented a model named VSUMM (Video SUMMarization),where static story board is generated based on modified k-means algorithm. Here the number of clusters is determined adaptively by a simple shot boundary detection method. Here the feature vector is generated only based on color histogram which makes this process too prone to spatiality problem. Also this process is computationally expensive for long length video. K. Kuanar *et al.* [4] has generated video summary based on improved Delaunay clustering and information-theoretic pre-sampling. A dynamic edge pruning method is applied to the Delaunay graph with some structural constraint to generate clusters. Both color and texture features used to generate hybrid feature vector, which more specifically identify content of a video. But this method does not ease user to provide their choice of specifying the number of key-frames to create a static video summary. Our proposed method is designed to remove some of the limitation of traditional clustering based methods. At the same time like [15] we maintain to provide the opportunity to the user to specify how many frames would be there in the static story board. This will generate the story board which will be more users friendly. To identify the content of a frame more specifically and to reduce spatiality problem we have used both color and texture features. As video summarization suffers from non-existence of standard evaluation framework, we have opted to use one objective metric (Compression Ratio (CR)) as used in [4] and a metric (Comparison of User Summaries (CUS)) used in [3], which we call semi-objective metric for summary evaluation.

# 3 Proposed Methodology

In Fig. 1, we have given an outline of the steps of our proposed method. It consists of mainly five steps: (1) sampling of video frames; (2) noise reduction; (3) feature extraction; (4) graph-based ranking; (5) key-frame extraction. Each component of our proposed system is detailed in the subsequent sections.

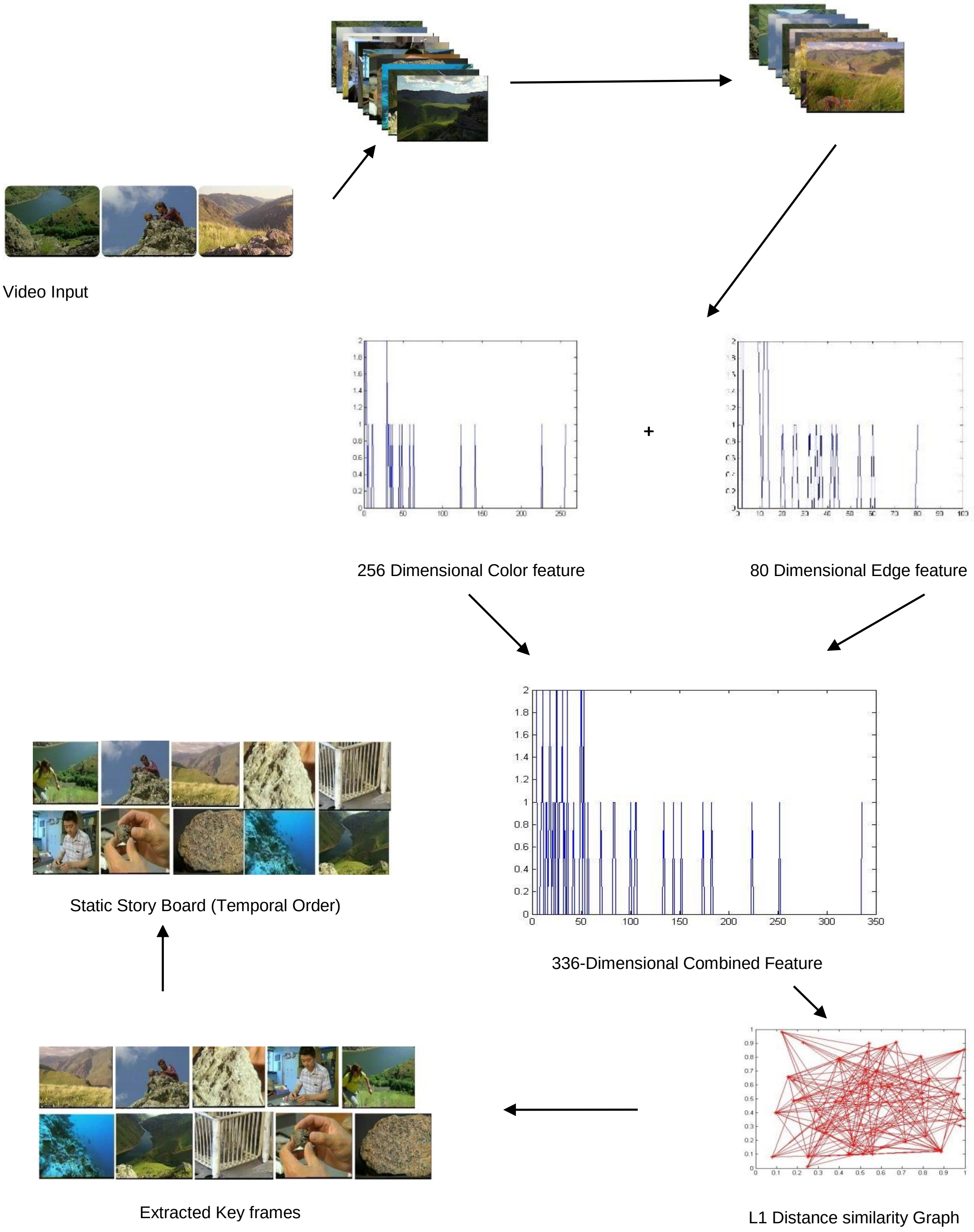

**Fig 1** Flowchart of our proposed VidRank approach

## 3.1 Sampling of video frames

Generally, in literature shot boundary detection is performed before video summarization. Another type of video segmentation is extraction of frames from video without considering the temporal order [3]. We have followed this method. Consecutive frames within a fraction of second have high probability of having almost similar content. So to reduce redundancy we have not considered all the frames. A fixed rate (1fps [3]) sampling method is applied on the original video. But the choice of sampling rate is very important because a high sampling rate can produce a low quality summary where a low sampling rate can lead to loss of information. We have followed one frame per second sampling rate [3].

## 3.2 Noise reduction

In our research , we have observed mainly one kind of noisy (i.e. *less meaningful*) frame, i.e. monochromatic frames. Processing with this type of frames will increase computational time, and also this type of frames can lead to produce low quality summary. So we have omitted these types of frames in pre-processing step i.e. before constructing the ranked graph. A property of monochromatic frame is that it should have high normalized variance between histogram bins because it follows homogeneous distribution [8]. We can see from Fig. 2 that normal frame has much lower variance than fade-in or fadeout frame. We have calculated mean and standard deviation of variance vector of all the frames of a video. Then the threshold (*T*) calculated adaptively as:

$$T = Mean + \beta * Standard\ Deviation \tag{1}$$

Where β is user defined positive constant. For our experiment we have chosen 1.8 as a value of β. If the variance an image of a video is above that threshold (*T*), then it should be treated as monochromatic junk frame and it is discarded.

## 3.3 Feature extraction

To identify an image uniquely from a large database feature plays a vital role. Choosing a correct feature highly dependent on what type of video we are going to work. We have chosen both color and texture feature to represent content of a frame. Each frame is represented by hybrid feature vector in multi-dimensional feature space.

### 3.3.1 Color Feature Extraction

Color is the most basic quality of visual content. We have used color histogram descriptor to represent color feature. Color histogram is a global descriptor of an image. In case of any type of linear transformation (e.g. rotation, translation) of an image, histogram of that image does not change so much. It is also robust in case of change of camera position. So, color histogram mostly resembles human perceptual behavior. Also HSV color model reflects the human perception of colors. Each frame is represented by 256 dimensional feature vector in HSV color space. According to MPEG-7 generic color histogram descriptor H is allocated 16 ranges, S 4 ranges and V 4 ranges.

### 3.3.2 Texture Feature Extraction

We have used Edge Histogram Descriptor [9] to extract texture feature of the frame. Each frame sub-divided into 4×4 blocks. For each of this block, five different types of edges are considered; vertical, horizontal, $45^o$ diagonal , $135^o$ diagonal, and isotropic (non-orientation specific). We have recognized these five different types of edges by 3×3 Sobel filters instead of 2×2 filters as described in [9], because 3×3 Sobel filter perform better than 2×2 [10]. Fig. 3 shows the five different filters which we used for edge detection purpose. Hence each local edge histogram is 5 dimensional. Like [9] each image partitioned into 16 sub-images which result 80 dimensional edge histogram representation of each frame. Histogram suffers from spatial information problem. So to reduce this problem we have incorporated texture feature in the feature space. This achieves higher semantic dependency between different video frames [4]. We have combined texture feature vector with color histogram feature vector by serial feature fusion strategy [11]. Now each frame is represented by 336 - dimensional feature vector.

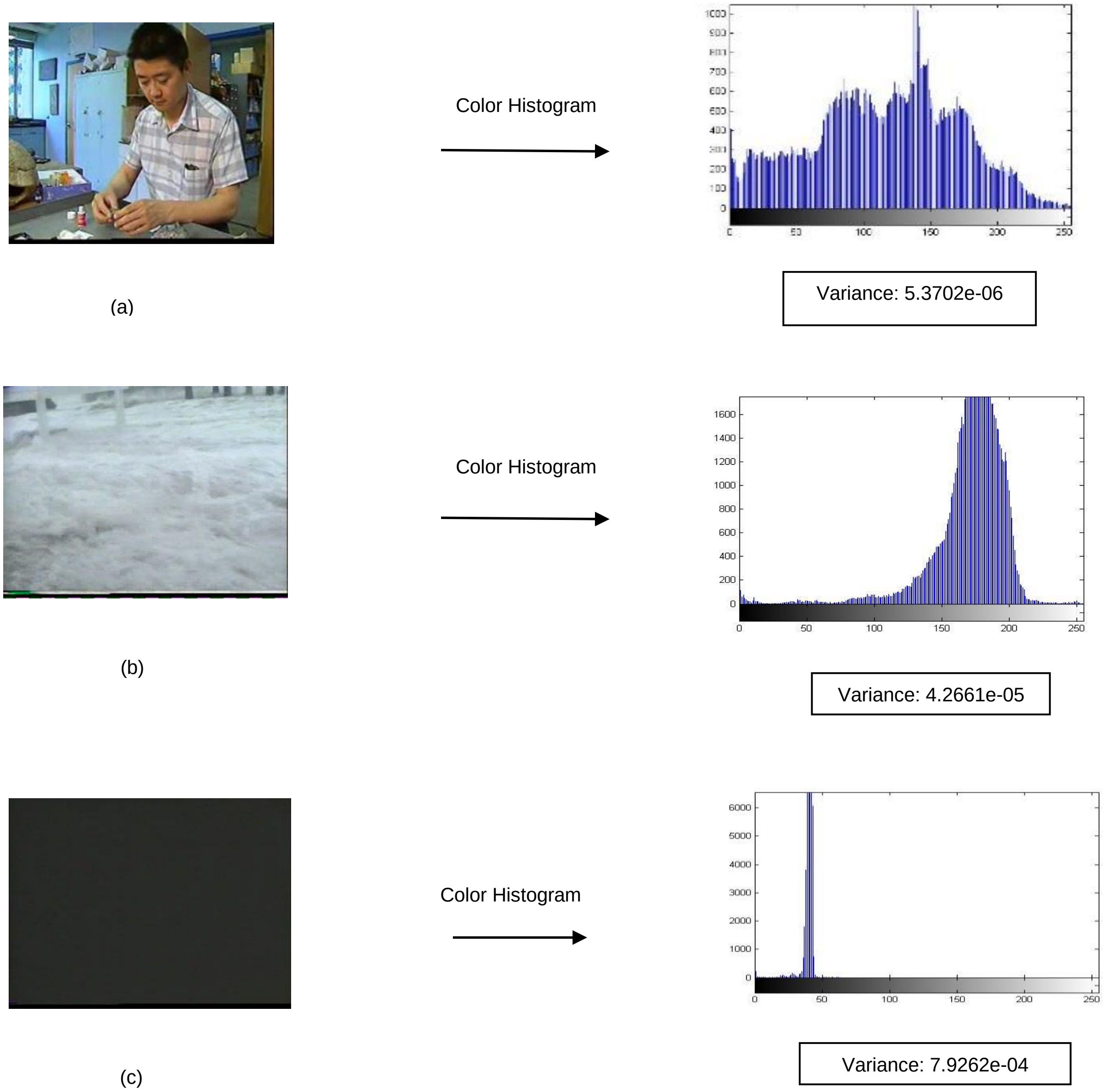

**Fig 2** Color Histogram and Variance of different frames (a) Normal frame, (b) fade-out frame, (c) fade-in frame

## 3.4 Graph-based Ranking

The main idea of graph-based ranking used in this paper is that, given a graph, we have to measure the "importance" or find "rank" of each vertex and a vertex which carries highest "importance" globally throughout the entire graph will be assigned the highest score (i.e. "importance"). We have treated each frame of a video as a vertex; there should be an edge between two vertices if both frames have a similarity above a threshold. We have used same formula (1) as mentioned in sub-section 3.2, to calculate adaptive threshold. We have used Manhattan Distance (i.e. L1 distance) for measuring dissimilarity between two frames because it performs best in CBIR system [12]. The graph thus constructed is an undirected graph. We have employed Random Walk intuition [18] to rank each vertex. The intuition is that if a user visits an image then other similar image is also of interest of that user i.e. if an image u has a visual similarity with another image v then there is some probability that the user will jump from u to v. Another intuition is that an image linked to a highly "important" image, should also carry high importance. We have named our approach as *VidRank*. *VidRank* can be defined as:

$$VDR = X * VDR \tag{2}$$

Here **X** is row normalized adjacency matrix representing the similarity graph. Initially we have set rank of each vertex equal to 1. *VidRank* converges only when **X** is aperiodic and irreducible. For introducing this property we have incorporated a damping factor d in (2) to obtain:

$$VDR = \left[ \left(1 - d\right) + dX \right] * VDR \tag{3}$$

Equation (3) mostly resemble "page rank" algorithm [2]. We have chosen value of d equal to 0.85 as mentioned in [2].

| | | |
|---|---|---|
| -1 | -2 | -1 |
| 0 | 0 | 0 |
| 1 | 2 | 1 |

Horizontal Mask

| | | |
|---|---|---|
| -1 | 0 | 1 |
| -2 | 0 | 2 |
| -1 | 0 | 1 |

Vertical Mask

| | | |
|---|---|---|
| -2 | -1 | 0 |
| -1 | 0 | 1 |
| 0 | 1 | 2 |

45 Degree Edge Mask

| | | |
|---|---|---|
| 0 | 1 | 2 |
| -1 | 0 | 1 |
| -2 | -1 | 0 |

135 Degree Edge Mask

| | | |
|---|---|---|
| -1 | 0 | 1 |
| 0 | 0 | 0 |
| 1 | 0 | -1 |

Non-directional Edge Mask

**Fig 3** Filters for edge detection

In our implementation we have augmented the basic page rank algorithm. We have evolved three different models of our implementation. In model 1, after getting rank value of each frame we have extracted highest ranked frame and then penalized its link vertices (i.e. vertices which are similar to it) by the following formula:

$$VDR(u) = VDR(u) - \alpha * SimValue(u,h) * VDR(h) \quad (4)$$

h refers to highest ranked vertex in a particular iteration, u refers to one of the vertices to be penalized due to their similarity with h (according to predefined threshold value), *SimValue ()* is a function which calculate similarity value between u and h, α is a user defined positive constant value, it is actually controlling parameter. For our experiment we have set α equal to 0.5. By penalizing method, we reduce the chance of choosing similar frame in static story board. The highest ranked frame of previous iteration is discouraged to be selected again in subsequent iteration. In model 2, we have modified (4):

$$VDR(u) = VDR(u) - \alpha * VDR(h) \quad (5)$$

In model 2 we uniformly penalize vertices by omitting SimValue () function. In model 3, we do not penalize frames, instead we have marked the similar frames and does not consider them for subsequent extraction of key-frames i.e. we delete similar vertices from the graph in each iteration.

$$VDR(u) = 0 \quad (6)$$

The iteration terminate either when the specified number of key frames have been extracted or the rank of all vertexes become zero or negative.

### 3.5 Key-frame Extraction

In spite of screen space problem associated with static story board display it is still preferred method over dynamic story board [16].In each iteration of *VidRank* we get a highest ranked frame which we consider as the most informative frame among its similar type of frames. So, we consider it as a key-frame which will construct the story board. Finally, the key frames are arranged in temporal order to make the video summary more understandable to the user.

## 4 EXPERIMENTAL RESULTS

In this section, we have compared our proposed video summarization models with three well known approaches presented in the literature [3-5]. In addition some information about our evaluation framework and experimental dataset is also provided.

### 4.1 Evaluation framework

A consistent evaluation framework is missing in video abstraction research. So every work has its own evaluation method. Due to lack of objective ground-truth video abstract is not a straightforward task. Even for a human it is difficult to decide whether a video abstract is better than another or not. For evaluation purpose we have used one objective, one semi-objective metric. For objective measure we used Compression Ratio (CR) [4]. Comparison of User Summaries (CUS) metric [3] is used as semi-objective measure. For semi-objective measurement purpose we have asked 5 different users to provide their summary observing each video of the video data set. Users have chosen number of key-frames in their own will. We have used same data set used in [3] collected from Open Video Database [14]. The experiment performed on three different genre videos.

**Compression Ratio (CR):** It measures the compactness of generated static story board. It defines how much concisely the entire story of a video can be represented. Compression Ratio (CR) of a video ($v$) can be defined as:

$$CR(v) = 1 - \left(\frac{M}{N}\right) \tag{7}$$

Where M is number of key frames and N is total number of frames of a video.

**Comparison of User Summaries (CUS):** In this method we collected summaries of users on same dataset used by our proposed models. Then we compare user summaries with summaries generated by our algorithm. For comparison purpose we have used color histogram intersection-based dissimilarity method [13]. Two frames are considered to be similar if there similarity is above a pre-defined threshold. Once two frames are matched they are removed for next iterations. Like [3] we have used two metrics of CUS, called accuracy rate CUS (A) and error rate CUS (E);

$$CUS(A) = \frac{N_m}{N_{us}} \tag{8}$$

$$CUS(E) = \frac{N_{nm}}{N_{us}} \tag{9}$$

Where $N_m$ denotes number of matched frames of system summary with respect to each user summary, $N_{nm}$ denotes number of non-matched frames of system summary with respect to each user summary and $N_{us}$ denotes number of key-frames from user summary.

## 4.2 Performance analysis

We tested our system on the data set described in [3]. All videos are MPEG-1 format (30 fps, 352 × 240 pixels). These videos are distributed in three genres i.e. documentary, lecture and educational. In documentary genre there are 44 numbers of videos, in educational genre there are 2 numbers of videos and in lecture genre there are 4 numbers of videos. We have performed our research work on Intel i-3, 1.80 GHz processor, 2 GB RAM.

Since our proposed graph-rank based video summarization method is less explored still now in video summarization field, we have opted cluster based video summarization methods for our performance comparison purpose because we have noticed that cluster based algorithm has some similarity with our proposed method. But for ideal comparison of video summaries of different approaches, it is needed that each one should be tested on same data set and evaluated using same metrics. Though the methods proposed in [15], [17] extracted fixed number of key frames like us; they have not made their data set available. So, we have used three mile-stone cluster based methods in literature [3-5] for our performance comparison purpose and also used same data set used by these methods. The summaries produced by the methods [3-5] are available at <http://www.sites.google.com/site/vsummsite/>.

In our rank-based system, we have to extract fixed number of frames for generating static story board. But the methods described in [3-5] have extracted variable number of key-frames. So, for meaningful systems comparisons, the number of key frames to be extracted is set to the average number of key-frames extracted by the methods [3-5], to which the proposed method is compared. We compared our proposed approach with the best models i.e. VSUMM (1) [3], DELAUNAY_CLUSTERING(C+E) [4] and with STIMO [5] reported in [3-5]. For comparison with VSUMM (1) and STIMO we have extracted 10 key-frames and for DELAUNAY_CLUSTERING (C+E) we have extracted 7 numbers of key-frames.

From Table 1 we can see, model 3 generates a comparable result. Though it suffers from low CUS (A) value and CUS (E) value, it produces more compact result than VSUMM (1) [3]. From Table 2 we can see, accuracy rate (CUS (A)) of model 3 is higher than DELAUNAY_CLUSTERING (C+E) [4] method and at the same time it also maintains same compactness of the summary. But for maintaining high divergence i.e. global description it tends to generate some other frames which are not matched to user summary, that's why it tends to generate little bit more CUS(E) value than DELAUNAY_CLUSTERING (C+E) [4]. It can be seen from Table 3 that our all three models produce better summery than STIMO [5] method. Though for model 1 and model 2 Compression Ratio is slightly low, its accuracy rate and error rate is better than STIMO [5]. Whereas model 3 produces same compact result as STIMO [5] and also maintains high accuracy rate and low error rate compared to other models. From Table 1, Table 2 and Table 3 it can be seen that VidRank (Model 3) out-performs both VidRank (Model 1) and VidRank (Model 2) in terms of both Compression Ratio, CUS(A) and CUS(E).So, among our three models we propose VidRank (Model 3) as our best model. We collected our own five different users review and performed all our evaluations and comparisons on that basis.

**Table 1** Comparison of our different models with VSUMM (1) [3] where number of key-frames to be extracted is set to 10 key-frames (which is equal to average number of key-frames extracted by VSUMM (1) [3])

| Metric Type | Measure | VSUMM(1) | VidRank (Model 1) | VidRank (Model 2) | VidRank (Model 3) |
|---|---|---|---|---|---|
| Semi- Objective | CUS(A) | 0.6789 | 0.6213 | 0.6213 | 0.6141 |
| | CUS(E) | 0.6785 | 0.8063 | 0.8063 | 0.7572 |
| Objective | Compression Ratio (CR) | 0.89 | 0.89 | 0.89 | 0.90 |

**Table 2** Comparison of our different models with DELAUNAY_CLUSTERING (C+E) [4] where number of key-frames to be extracted is set to 7 key-frames (which is equal to average number of key-frames extracted by DELAUNAY_CLUSTERING (C+E) [4])

| Metric Type | Measure | DELAUNAY_CLUSTERING (C+E) | VidRank (Model 1) | VidRank (Model 2) | VidRank (Model 3) |
|---|---|---|---|---|---|
| Semi- Objective | CUS(A) | 0.4551 | 0.4811 | 0.4829 | 0.4830 |
| | CUS(E) | 0.4947 | 0.5789 | 0.5712 | 0.5656 |
| Objective | Compression Ratio (CR) | 0.93 | 0.92 | 0.92 | 0.93 |

**Table 3** Comparison of our different models with STIMO [5] where number of key-frames to be extracted is set to 10 key-frames (which is equal to average number of key-frames extracted by STIMO [5])

| Metric Type | Measure | STIMO | VidRank (Model 1) | VidRank (Model 2) | VidRank (Model 3) |
|---|---|---|---|---|---|
| Semi- Objective | CUS(A) | 0.5246 | 0.6213 | 0.6213 | 0.6141 |
| | CUS(E) | 0.9019 | 0.8063 | 0.8063 | 0.7572 |
| Objective | Compression Ratio (CR) | 0.90 | 0.89 | 0.89 | 0.90 |

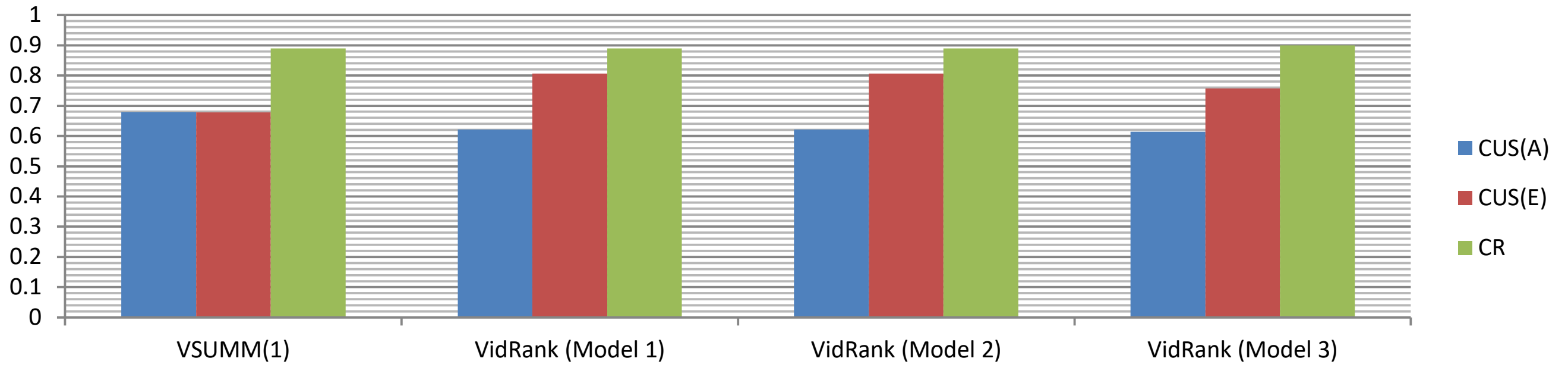

**Fig 4** Comparison of our propose three models with VSUMM (1) [3]

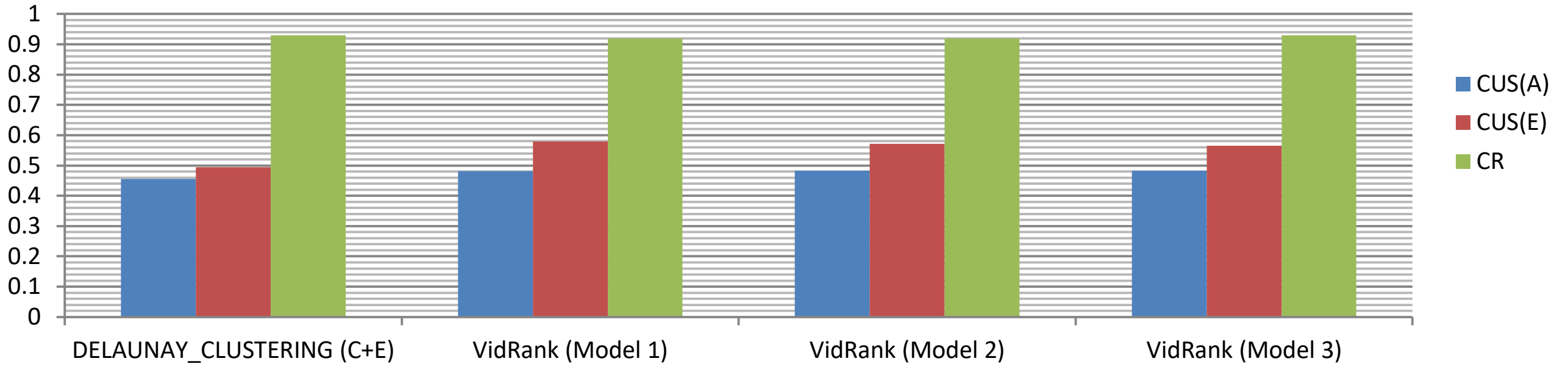

**Fig 5** Comparison of our propose three models with DELAUNAY_CLUSTERING (C+E) [4]

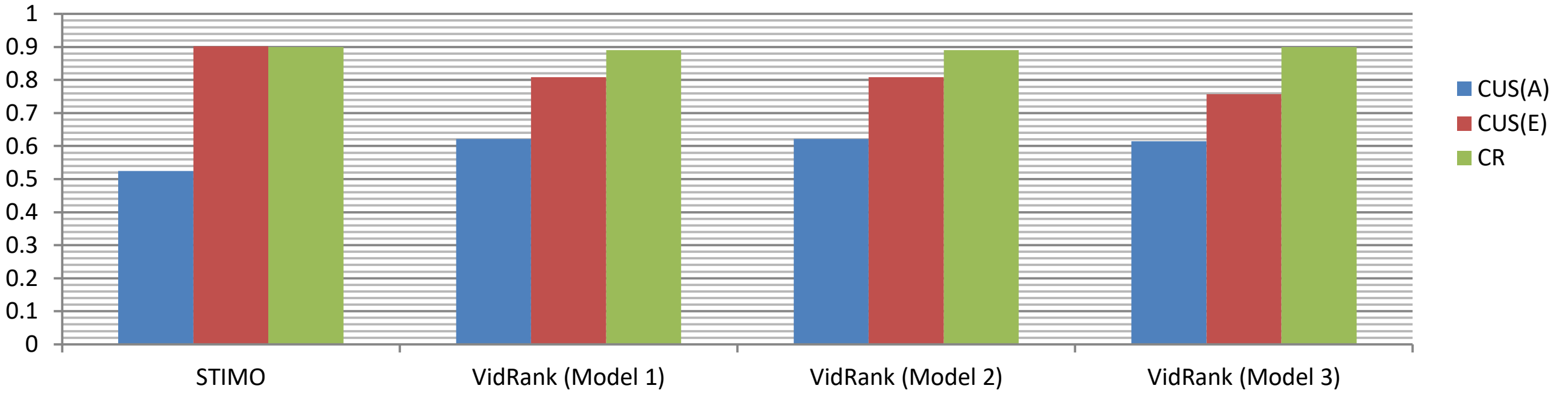

**Fig 6** Comparison of our propose three models with STIMO [5]

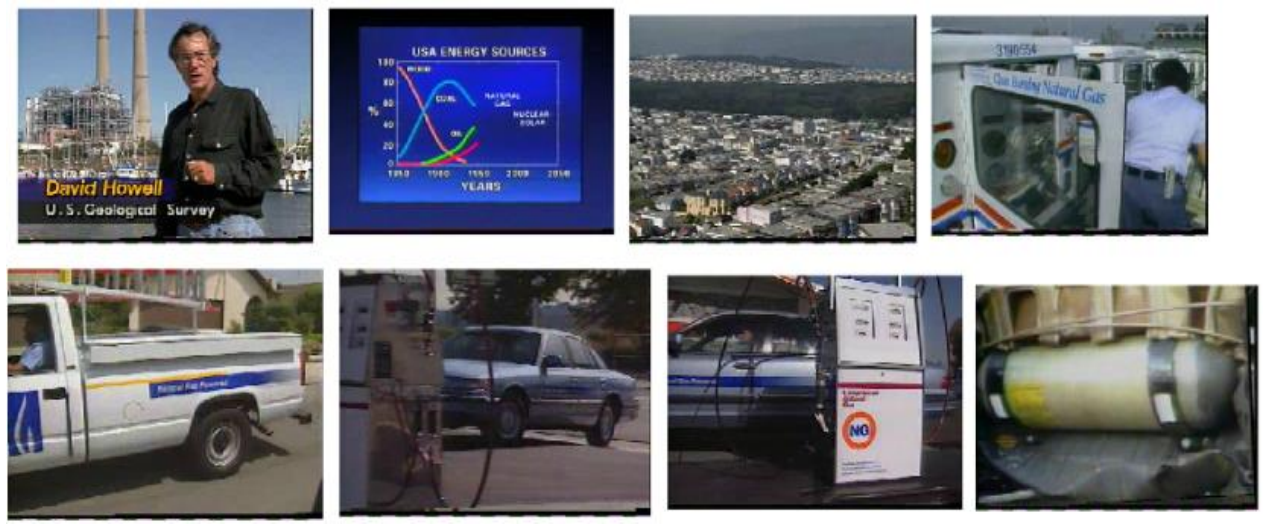

(a) VSUMM(1)

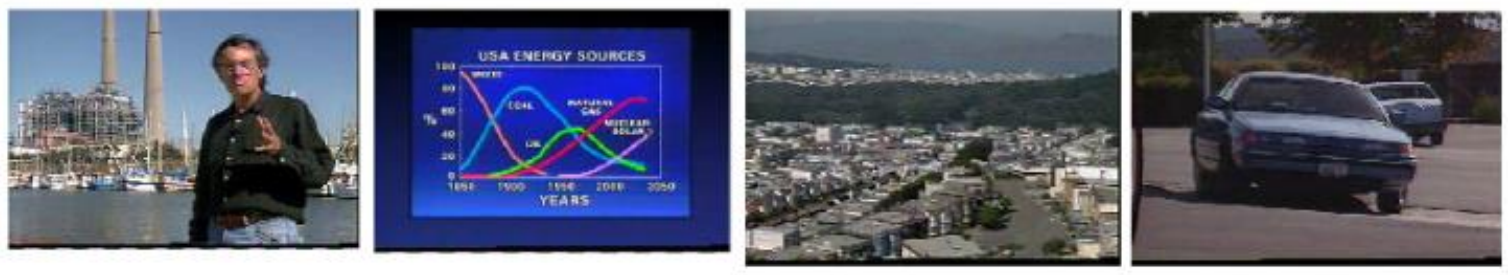

(b) DELAUNAY_CLUSTERING (C+E)

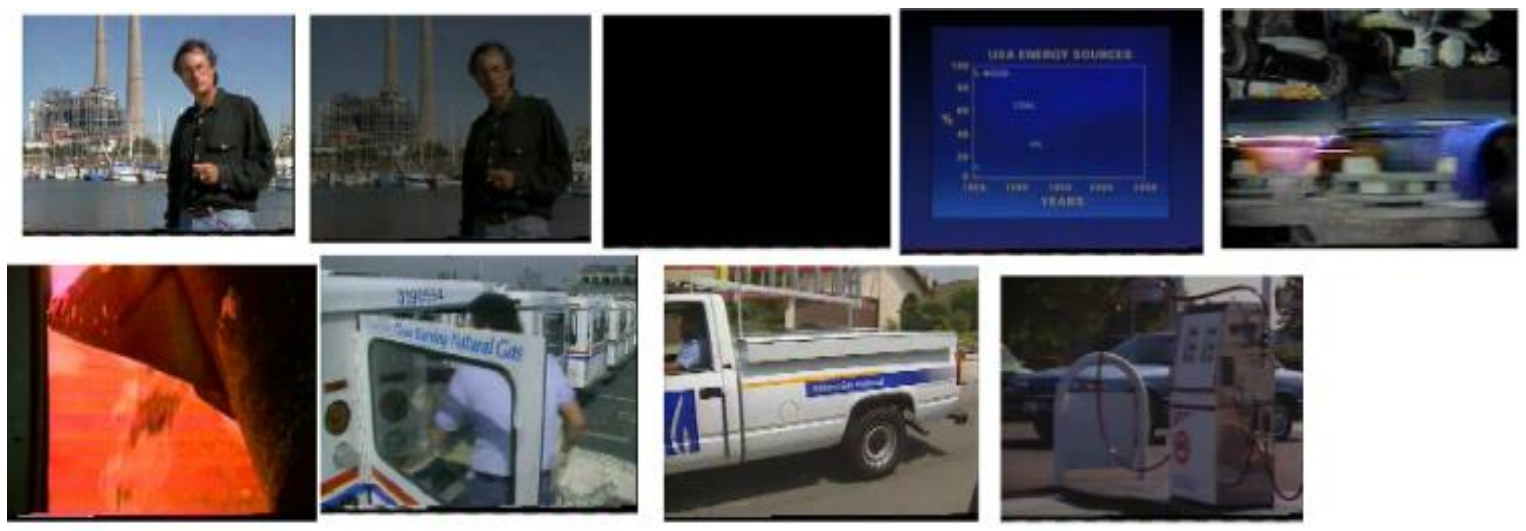

(c) STIMO

**Fig 7** Video summaries of different approaches: (a) VSUMM (1) (b) DELAUNAY_CLUSTERING (C+E) and (c) STIMO of the video on natural gas (available at open video project [14])

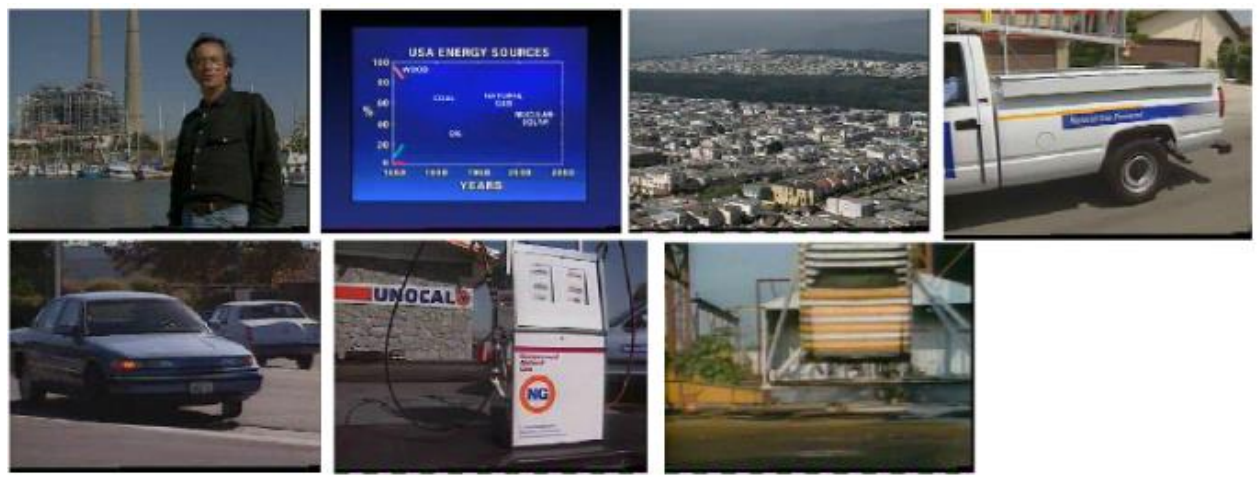

(a)

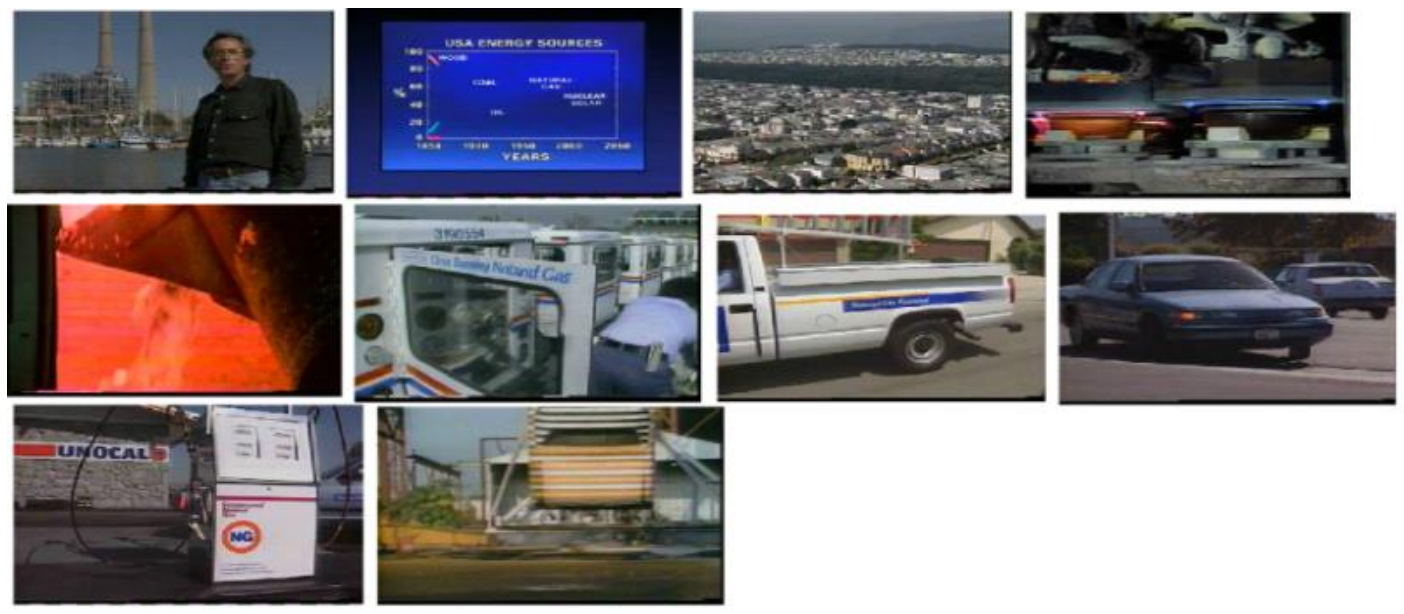

(b)

**Fig 8** Video summary produced by VidRank (Model 1) of the video on natural gas; (a) for extraction of 7 key-frames, (b) for extraction of 10 key-frames

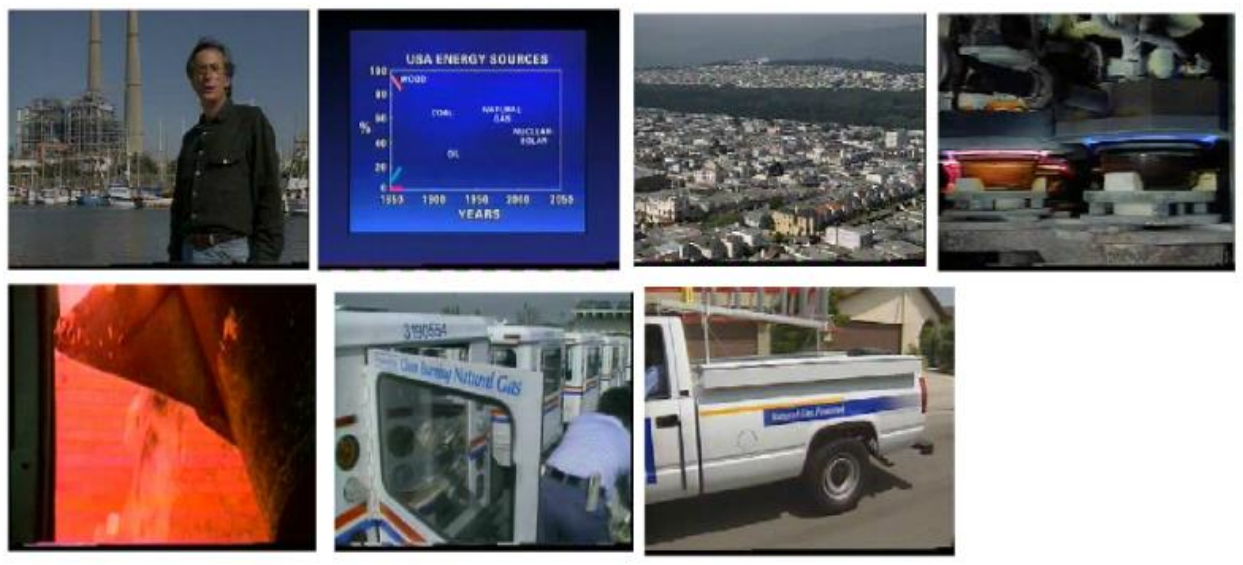

(a)

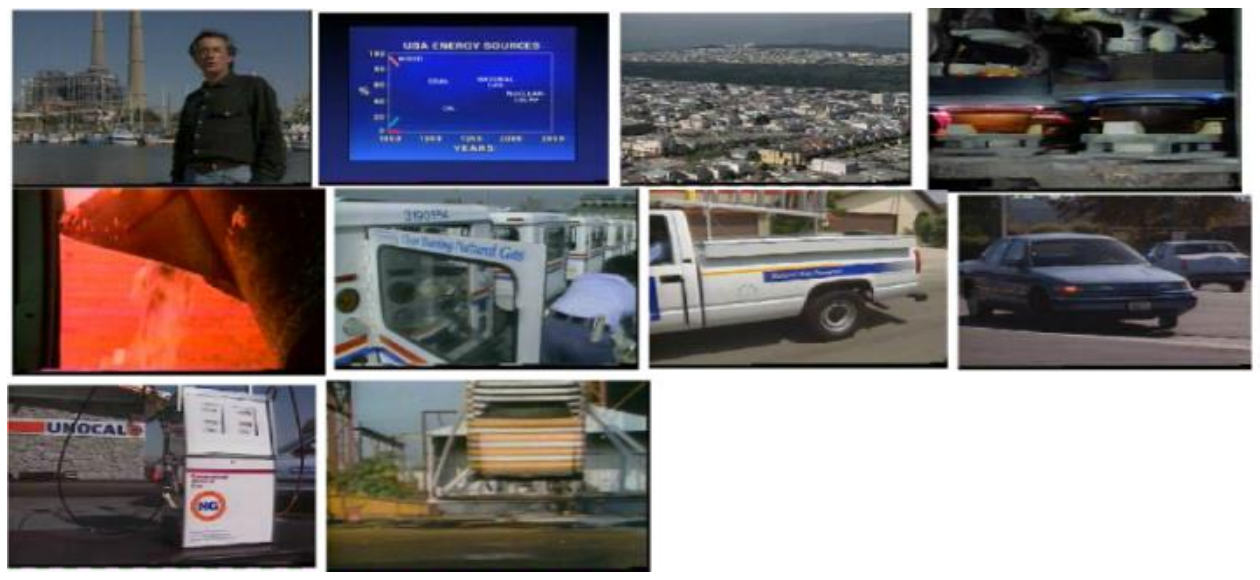

(b)

**Fig 9** Video summary produced by VidRank (Model 2) of the video on natural gas; (a) for extraction of 7 key-frames, (b) for extraction of 10 key-frames

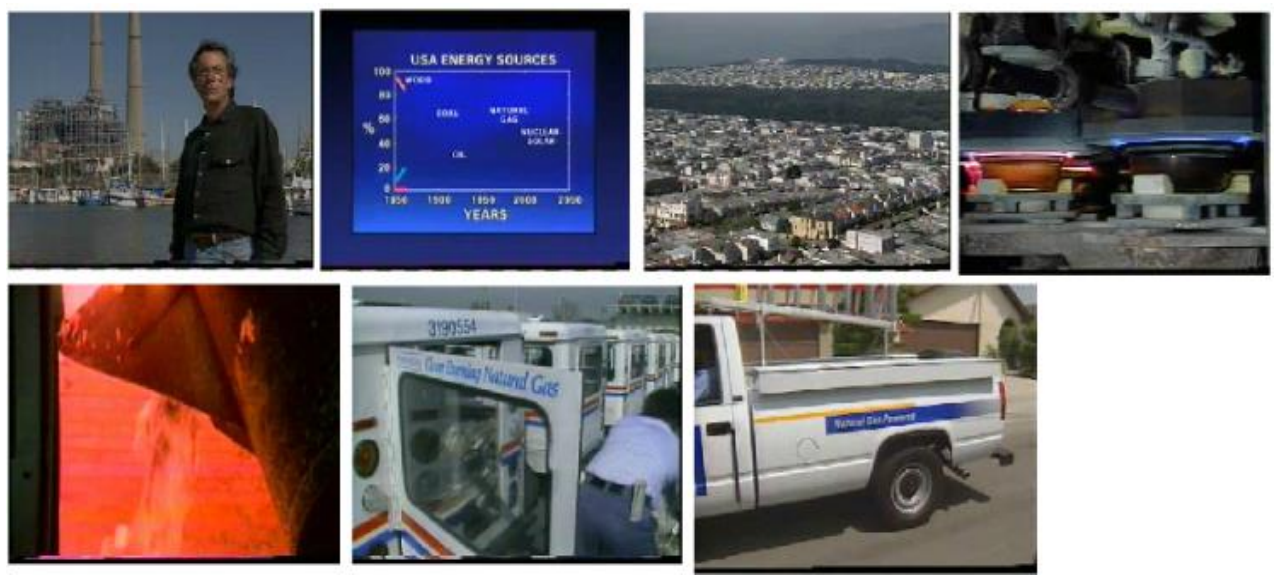

(a)

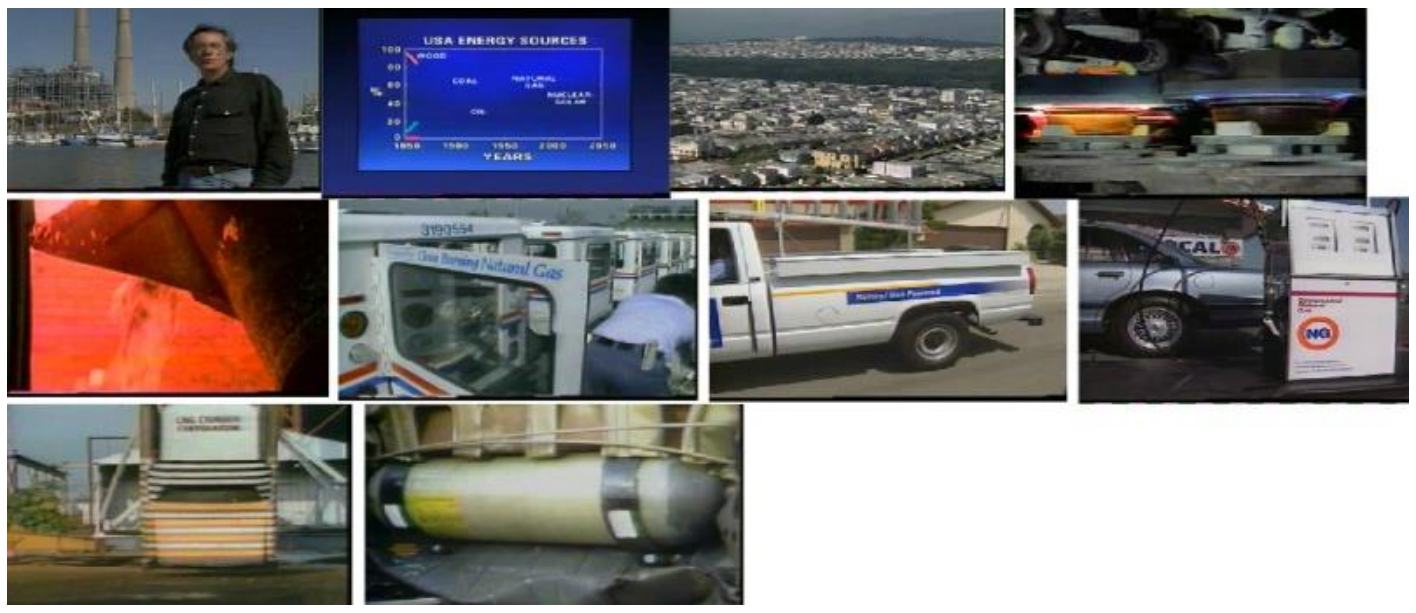

(b)

**Fig 10** Video summary produced by VidRank (Model 3) of the video on natural gas; (a) for extraction of 7 key-frames, (b) for extraction of 10 key-frames

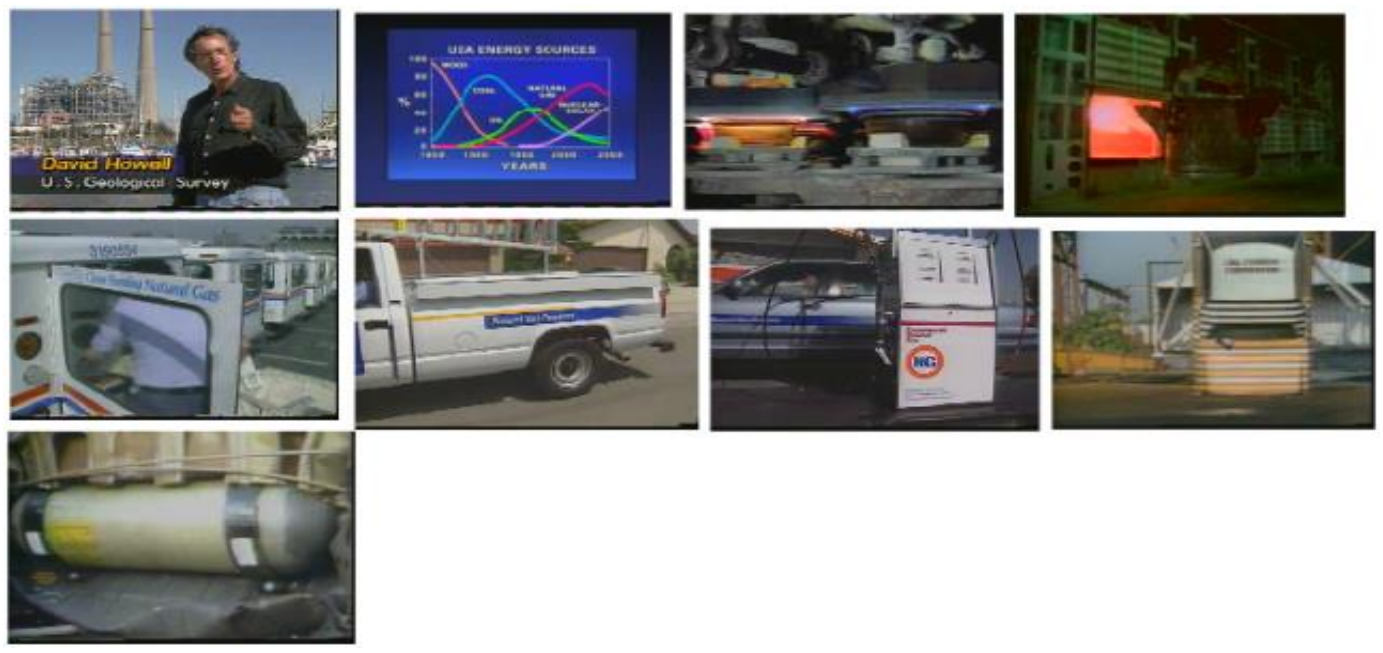

(a) User 1

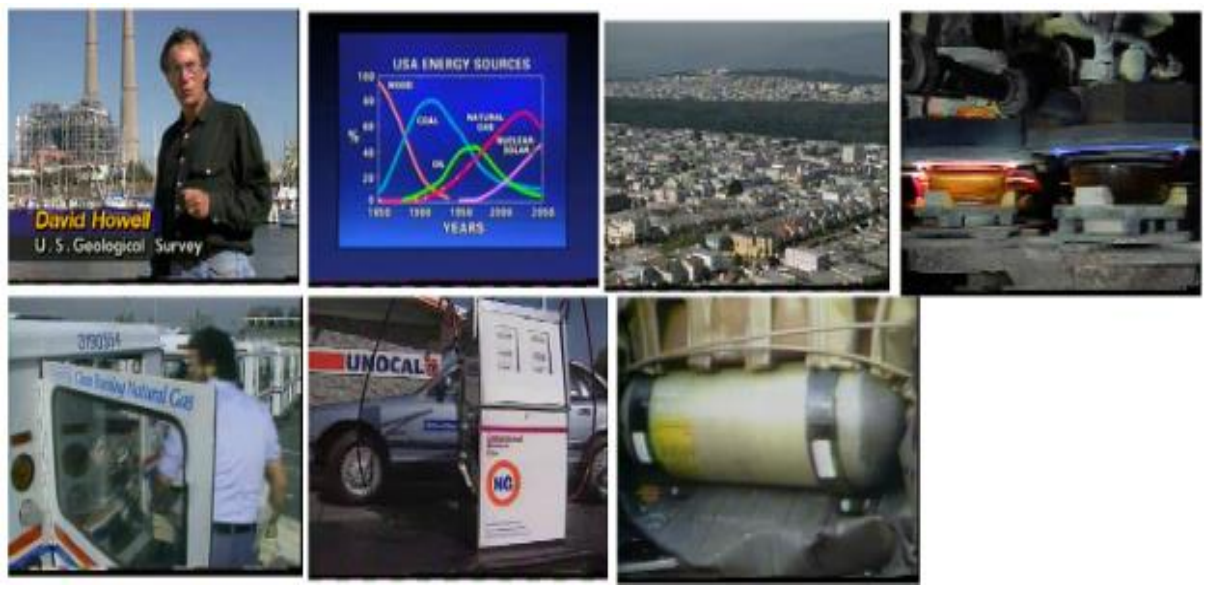

(b) User 2

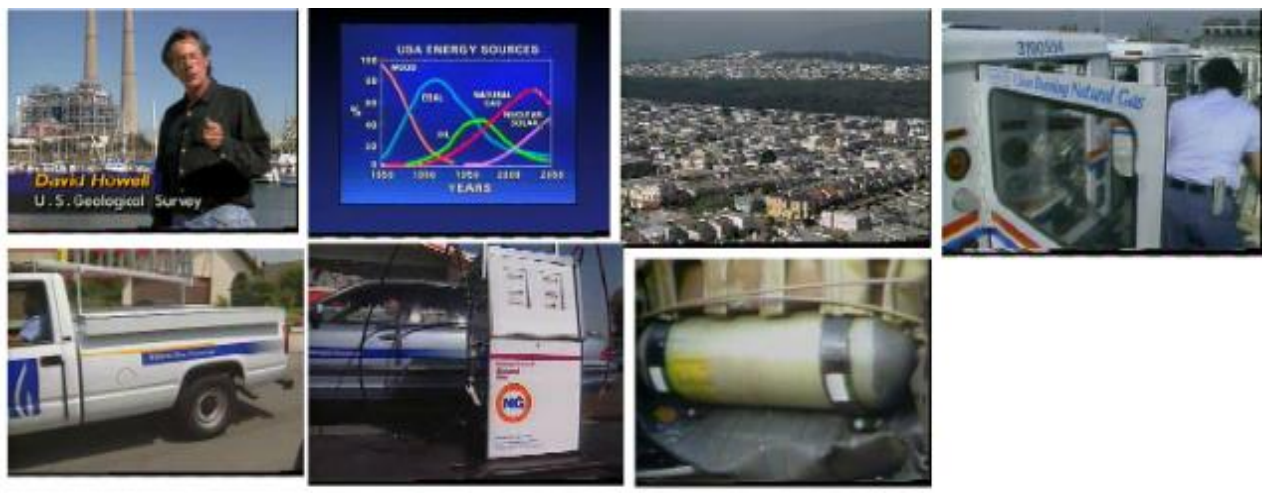

(c) User 3

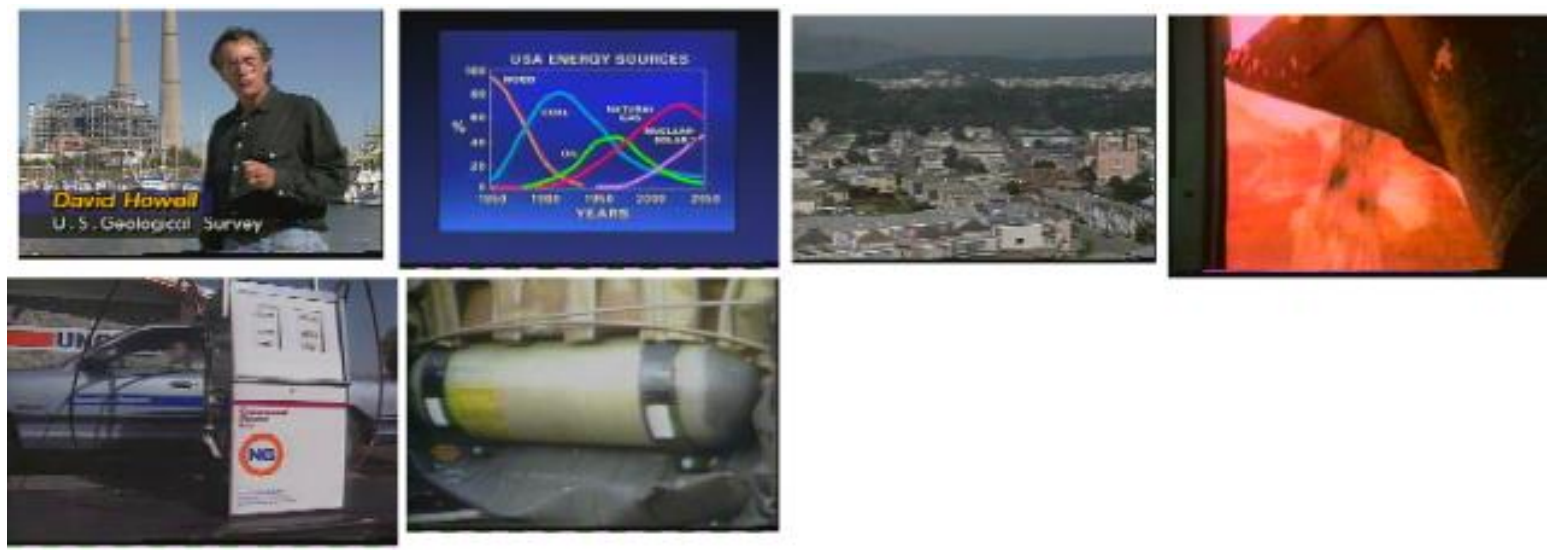

(d) User 4

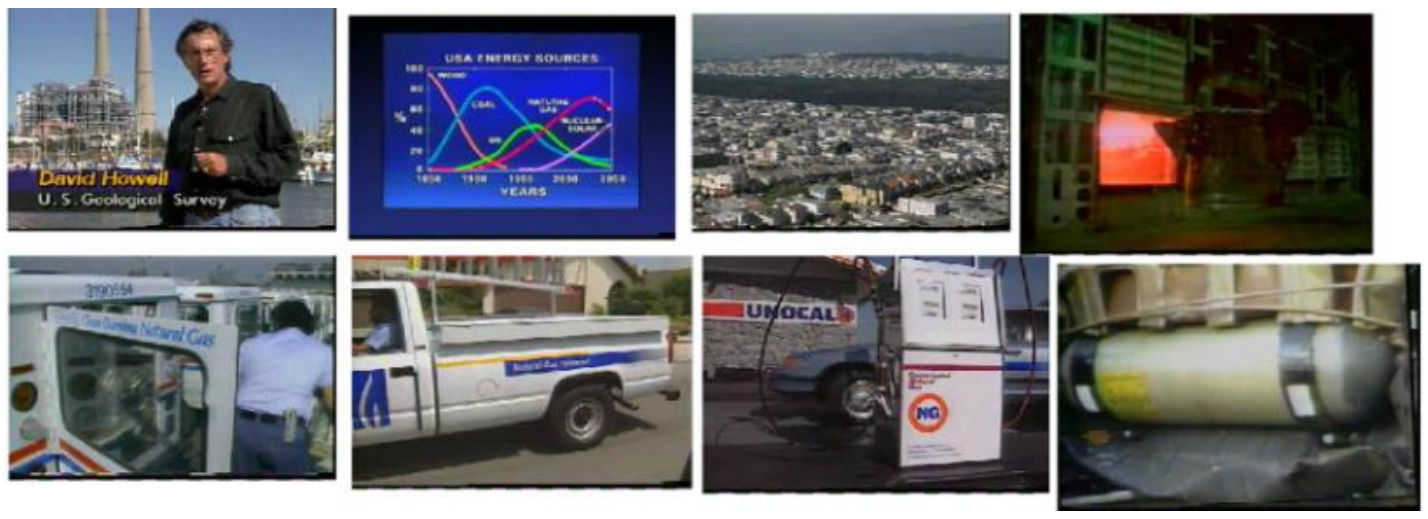

(e) User 5

**Fig 11** User summaries of the video on natural gas

Fig 4, 5, and 6 shows graphical representation of our experimental comparison results. Fig 7 shows video summaries produced by different approaches (i.e. VSUMM (1), DELAUNAY_CLUSTERING (C+E) and STIMO) on the video of natural gas. Fig 8, 9 and 10 shows video summary produced by our proposed model 1, model 2 and model 3 on the video of natural gas. From Fig 10 and Fig 11 it can be seen that our proposed model 3 mostly resembles with user generated summaries.

# 5 Conclusions and Future Work

In this paper, we proposed an approach to generate static story board, which is less explored in video summarization domain. We have shown that our graph-based approach mostly resembles the page-rank based approach. We have made three different models of summarization based on a hybrid feature vector. Due to absence of standard evaluation framework we have adopted a combined evaluation metric framework consisting of one purely objective metric (Compression Ratio) and another metric which have both subjective and objective quality. We termed this as semi-objective metric. Both in algorithm and in evaluation part we have chosen adaptive threshold instead of rigid threshold. We have shown that our proposed model perform well to generate static story board. In comparison with the existing systems in static video summarization domain the output of this model is encouraging.

Though this work is comparable to some best known summarization systems, there is still lot of scope to increase the accuracy further. Firstly, it is possible to add some more features in feature space which should able this algorithm to more uniquely identify frames. Secondly, the present work done its evaluation only on three different genre video, it can be tested further for other type of videos also, which should measure its scalability also. Though we have considered fade-in, fade-out type transition and successfully removed it, but we did not consider to remove dissolve and wipe type transition. In future some other robust noise removal techniques may be adopted to produce better quality results.